\documentclass[aps,prl,a4paper,twocolumn,superscriptaddress,showkeys,longbibliography]{revtex4-1}

\usepackage[colorlinks=true,bookmarks=false,citecolor=blue,linkcolor=red,hyperfootnotes=true,urlcolor=blue]{hyperref}

\usepackage{physics}
\usepackage{graphicx}
\usepackage{dcolumn}
\usepackage{bm}
\usepackage[dvipsnames]{xcolor}
\usepackage{wasysym}
\usepackage{lipsum}
\usepackage{enumitem}
\usepackage{booktabs} 
\usepackage{gensymb}
\usepackage{xcolor}
\usepackage{amssymb}
\usepackage{booktabs}
\usepackage{array}
\newcolumntype{L}{>{$}l<{$}}
\newcolumntype{C}{>{$}c<{$}}
\newcolumntype{R}{>{$}r<{$}}

\usepackage{mathtools}
\usepackage{siunitx}
\usepackage{soul}
\usepackage{float}

\makeatletter
\def\p@subsection{}
\makeatother

\makeatletter
\def\maketitle{
\@author@finish
\title@column\titleblock@produce
\suppressfloats[t]}
\makeatother

\newcommand{\hatc}{\hat{c}}

\begin{document}

\title{Strongly correlated ladders in K-doped $p$-terphenyl crystals}

\author{John Sous} \email{js5530@columbia.edu}
\affiliation{Department of Physics, Columbia University, New York,
New York 10027, USA}

\author{Natalia A. Gadjieva} 
\affiliation{Department of Chemistry, Columbia University, New York,
New York 10027, USA}

\author{Colin Nuckolls} 
\affiliation{Department of Chemistry, Columbia University, New York,
New York 10027, USA}
\author{David R. Reichman} \email{drr2103@columbia.edu}
\affiliation{Department of Chemistry, Columbia University, New York,
New York 10027, USA}

\author{Andrew J. Millis} \email{ajm2021@columbia.edu}
\affiliation{Department of Physics, Columbia University, New York,
New York 10027, USA} 
\affiliation{Center for Computational Quantum Physics, Flatiron Institute, 162 5$^{th}$ Avenue, New York, NY 10010}

\date{\today}

\begin{abstract}
Potassium-doped terphenyl has recently attracted attention as a potential host for high-transition-temperature superconductivity.  Here, we elucidate the many-body electronic structure of recently synthesized potassium-doped terphenyl crystals. We show that this system may be understood as a set of weakly coupled one-dimensional ladders. Depending on the strength of the inter-ladder coupling the system may exhibit spin-gapped valence-bond solid or antiferromagnetic phases, both of which upon hole doping may give rise to superconductivity. This terphenyl-based ladder material serves as a new platform for investigating  the fate of ladder phases  in presence of three-dimensional coupling as well as for novel superconductivity.
\end{abstract}

\maketitle

\emph{Introduction}.--- 
Recent reports~\cite{ChineseTerphenyl,ItalianTerphenyl} of  temperature-dependent gaps suggestive of superconductivity with a transition temperature of up to $\sim$ 100 K in powders formed from annealed alkali-doped terphenyl molecules have generated both excitement and questions. Terphenyl molecules are organic compounds based on linked aromatic (benzene) rings that exhibit strong electronic correlations, but until recently could not be doped to levels relevant for strongly correlated electronic and magnetic phases. The materials on which the recent reports are based are however annealed powders  not single crystals, which, from a synthetic and structural point of view, cannot be fully well characterized~\cite{Decompose}. Progress requires convergence of experiment and theory based on well characterized materials with specific crystal structures. A recent in-solution synthetic approach presents an alternative path to the synthesis of doped $p$-terphenyl {\em crystals} with well characterized structures~\cite{ColumbiaTerphenyl}, opening a door to the  systematic study of correlation-driven physics in these systems~\footnote{Note that well characterized crystals of alkali-doped terphenyl were also reported in the 1980s~\cite{Oldterphenyl}.}.

\begin{figure}[!t]
\centering
\includegraphics[width=0.847\columnwidth]{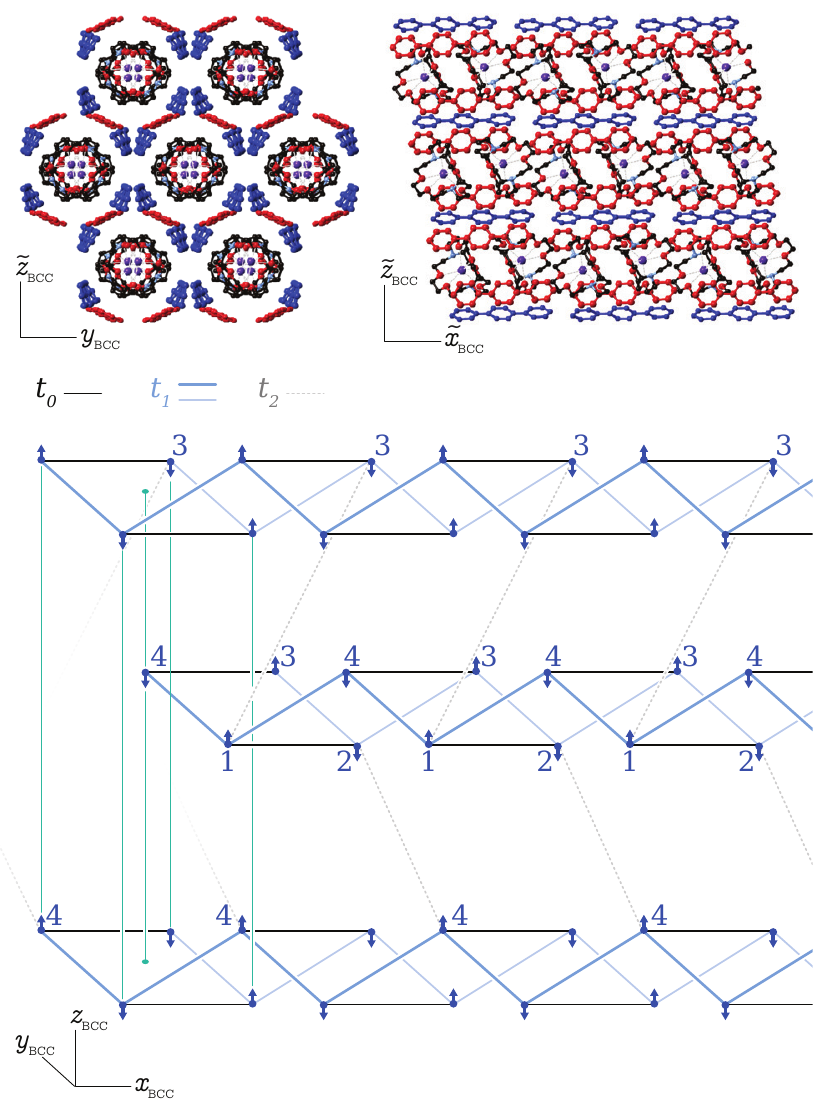}
\caption{{\bf Structure of [K($\mathbf{222}$)]$_2$[$p$-terphenyl]$_3$ crystal.} Upper panel: Crystal structure in two different orientations corresponding approximately (tilde) to planes defined by the body-centered cubic (BCC) lattice axes introduced in the text. In the left image, corresponding to the $y_{\rm BCC}$-$\tilde{z}_{\rm BCC}$ plane, note visible channels of the (222) (cage) chelated to four K$^+$ ions (violet dots), surrounded by $p$-terphenyl molecules that are either doped (blue) or neutral (red). In the right image, corresponding to the $\tilde{x}_{\rm BCC}$-$\tilde{z}_{\rm BCC}$ plane, note the chain pattern of terphenyl molecules along $\tilde{x}_{\rm BCC}$. Lower panel: Schematic representation of the tight-binding model corresponding to the  BCC lattice of coupled ladders with $t_0$ (black solid lines) and $t_1$ (heavy and light solid blue lines) hoppings within individual ladders and $t_2$ hopping (dotted gray lines) coupling different ladders. Note $t_1$ hopping within a ladder connects orbitals in different unit cells. We employ different thickness of the $t_1$ bonds as a guide to the eye. Thin vertical lines represent the edges  of the BCC lattice. 
}
\vspace{-4mm} 
\label{fig:Fig1}
\end{figure}

In this work, we present a theory of electronic structure and correlations in the potassium-doped terphenyl crystals reported in Ref.~\cite{ColumbiaTerphenyl}.  We find that this system can be accurately described in terms of a model of coupled ladders at half filling with an inter-ladder hopping amplitude about half as large as the intra-ladder hopping amplitudes.  This class of models exhibits a transition as a function of the inter-ladder coupling strength from a  spin-gapped paramagnet to a gapless antiferromagnetically ordered state~\cite{DagottaCoupledLadders1,RiceCoupledLadders1}.  Our computed value of the inter-ladder coupling places this material near to the transition on the antiferromagnetic side of the phase diagram. Using literature results on related models, we sketch the physical behavior, including potential $d$-wave superconductivity, expected upon doping the parent compound with holes and propose experiments. Our results establish terphenyl-based supramolecular materials whose structure can be controllably manipulated synthetically as a new family of ladder materials within which to explore correlation-driven emergent electronic and magnetic behavior.

{\em Crystal structure}.--- The recently synthesized crystals of potassium-doped para-terphenyl,  [K($\mathbf{222}$)]$_2$[$p$-terphenyl]$_3$~\cite{ColumbiaTerphenyl}, consists of two potassium (K) atoms, each within a cryptand ($\mathbf{222}$), for every three $p$-terphenyl (C$_{18}$H$_{14}$) molecules, see Fig.~\ref{fig:Fig1}). The cryptand is a polydentate ligand formed of nitrogen, oxygen, carbon and hydrogen atoms used in a solution-based approach to hold the dopant K atoms in place within the crystal structure upon crystallization. The synthesized crystals are monoclinic, with a primitive unit cell formed of two units of [K($\mathbf{222}$)]$_2$[C$_{18}$H$_{14}$]$_3$, containing four K atoms, six terphenyl molecules and a total of 444 atoms per unit cell. The cell parameters: $\vec{a} = \{0, 13.0237, 0.0\}$\si{\angstrom}, $\vec{b} = \{13.694074, 6.51185, 8.698852 \} $\si{\angstrom} and $\vec{c} = \{13.694074, 6.51185, -13.976448 \}$\si{\angstrom} in Cartesian $x,y,z$ coordinates.  Note $\vec{a}$ is purely along one of the principal axes of the system, which implies that a two-dimensional (2D) structure may be possible to engineer, for example by exfoliation or adsorption of a solution to a surface. We also note that this system differs significantly from  optimized structures obtained in simulations of doped $p$-terphenyl crystals without cryptand ligands~\cite{JPhysChemCDFT}. The cryptand leads to larger inter-terphenyl distances  than those simulated in Ref.~\cite{JPhysChemCDFT}, and enables a different level of doping. The compound we study corresponds to a level of doping of less than one K atom per terphenyl molecule (or of one K atom per \emph{active} terphenyl molecule, see below), while the structure studied in Ref.~\cite{JPhysChemCDFT} considers a range of doping levels from one to three K atoms per terphenyl molecule.

{\em Electronic structure}.--- We perform {\em ab-initio} density-functional theory (DFT) calculations using the Quantum ESPRESSO software package~\cite{QuantumESSPRESSO}, utilizing the Perdew, Burke and Ernzerhof (PBE) exchange correlation functional~\cite{PBE}, see Supplementary Information for details.  We note that prior spin-polarized DFT calculations revealed a stable three-dimensional (3D) antiferromagnet formed from the terphenyl molecules in this crystal structure~\cite{ColumbiaTerphenyl}. These calculations are in effect a Hartree-Fock approximation that cannot capture the spin-gapped state of isolated ladders and may in general overestimate magnetism. The ground state and exchange coupling constants were reported in this work but not the band structure. Here, we aim to characterize the system's spin-unconstrained electronic band structure and indicate how to incorporate electron-electron and electron-lattice correlations in order to understand correlation-driven behavior and emergent phases characteristic of this class of organic systems.

Figure~\ref{fig:Fig2} summarizes our main result concerning the electronic structure of this system. Our DFT band structure calculations (Fig.~\ref{fig:Fig2}, upper panel; left column) reveal six electronic bands near the Fermi level $\mathrm{E_{F}}$, with four lower-energy bands whose bandwidth ranges in the interval $\sim 50$ - $75 \mathrm{meV}$ roughly symmetric about $\mathrm{E_{F}}$, and two higher-energy bands whose bandwidth lies in the $\sim 10 \mathrm{meV}$ range located above $\mathrm{E_{F}}$. This doped crystalline structure exhibits much narrower bands than those found in the theoretically generated structures of Ref.~\cite{JPhysChemCDFT}, likely due to the extended space between terphenyl molecules occupied by the cryptand molecules in this crystal, absent in simulated optimized structures studied in Ref.~~\cite{JPhysChemCDFT}.
To understand the bands, we review the electronic structure of undoped terphenyl molecules.  Puschnig and Drexl~\cite{PuschingDrexel} used DFT calculations for single  oligophenyl molecules (e.g. terphenyl) isolated in space to argue for a H\"uckel molecular orbital picture in which the lowest unoccupied molecular orbital (LUMO) is localized on the molecule's aromatic rings. For the structure considered in this work, this means that the six terphenyl molecules in the unit cell contribute six states, giving rise to six bands near the Fermi level.  We also expect that each of the four dopant K atoms donates a single unit of charge to the terphenyl molecules. Single-crystal X-ray diffraction experiments reveal four equivalent terphenyl molecules with relatively shorter inter-phenyl bond lengths (shown in blue in Fig.~\ref{fig:Fig1}, upper panel) and two others with a relatively longer bond length (shown in red in Fig.~\ref{fig:Fig1}, upper panel) within each unit of [K($\mathbf{222}$)]$_4$[$p$-terphenyl]$_6$~\cite{ColumbiaTerphenyl}. We associate the four dispersive bands with the LUMO states of the four short-bond molecules and argue that these contain the four electrons donated by the K atoms.

 \begin{figure*}
 \centering
    \includegraphics[width=1.693\columnwidth]{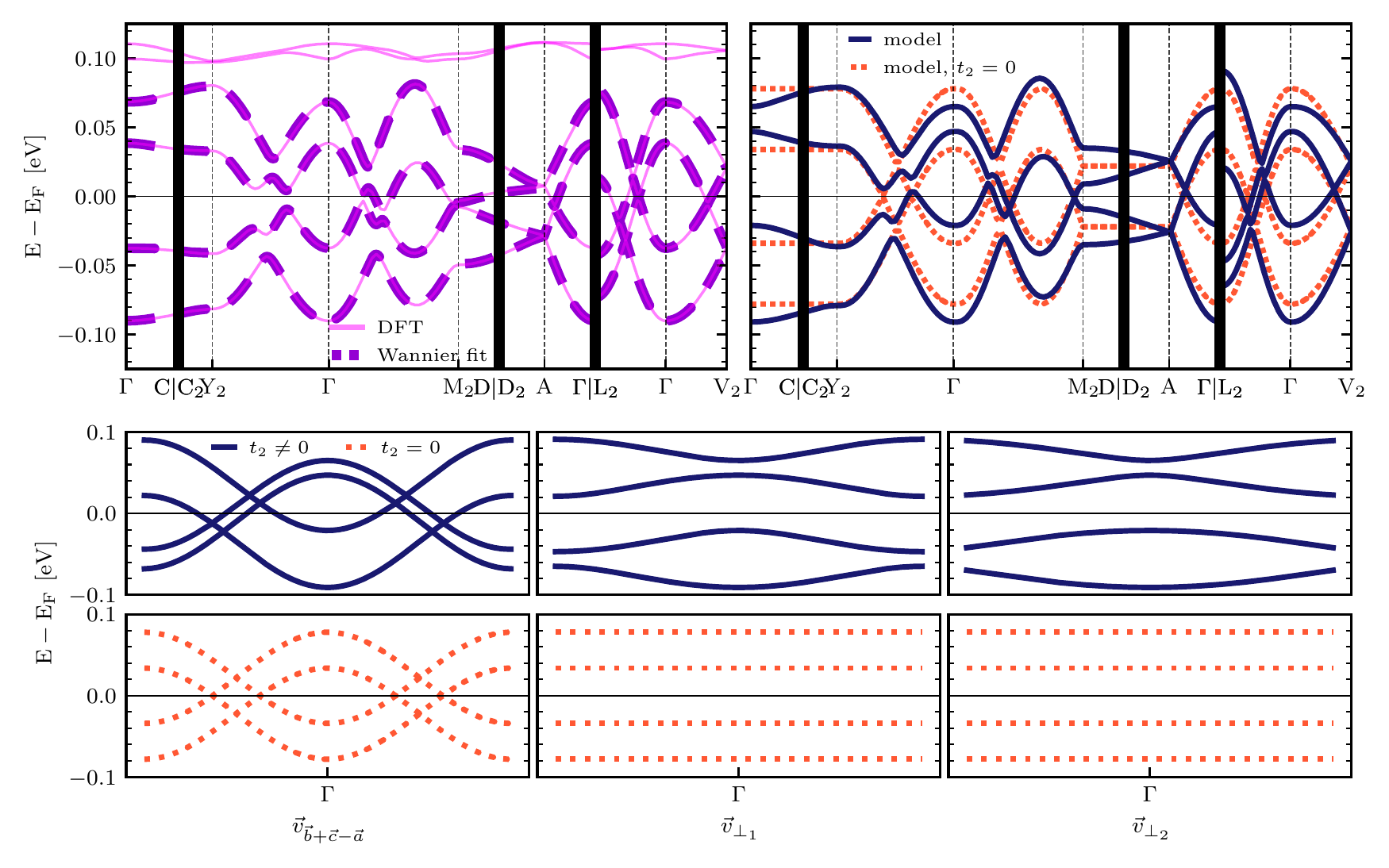}
    \caption{{\bf Electronic band structure of [K($\mathbf{222}$)]$_2$[$p$-terphenyl]$_3$ crystal.} Upper panel: Electronic band structure from DFT calculations and Wannier fits (left column) and from the minimal model with all Hamiltonian matrix elements smaller in magnitude than than 0.01eV neglected (right column) for $t_2 \neq 0$ and $t_2 = 0$. Lower panel: Electronic band dispersion from the minimal model along $\vec{v}_{\vec{b}+\vec{c}-\vec{a}} = (\vec{b}+\vec{c}-\vec{a})/|\vec{b}+\vec{c}-\vec{a}|^2$ (left) and its two orthogonal directions  $\vec{v}_{\perp_1} = \vec{a}/|\vec{a}|^2$ (center) and $\vec{v}_{\perp_2} = {\mathcal{R}_{x\rightarrow z, y\rightarrow y, z\rightarrow-x}} [\vec{v}_{\vec{b}+\vec{c}-\vec{a}}]$ (right) for $t_2 \neq 0$ (top) and $t_2 = 0$ (bottom). Here we interpolate the dispersion uniformly along $[-\vec{v},\vec{v}]$ through the $\Gamma$ point.  ${\mathcal{R}_{x\rightarrow z, z\rightarrow-x}} = \begin{pmatrix}
0 & 0 & -1\\
0 & 1 &  0\\
1 & 0 &  0
\end{pmatrix}$ denotes a rotation matrix that takes $x\rightarrow z$, $y\rightarrow y$, $z\rightarrow-x$.}
    \label{fig:Fig2}
\vspace{-4mm}     
\end{figure*}

{\em Minimal model}.---  We use  the wannier90 software package~\cite{Wannier90} to obtain a tight-binding representation of the near-$\mathrm{E_{F}}$ bands~\cite{WannierLocalization}, see Supplementary Information for details. We obtain Wannier fits of six- (not shown) and four-band (Fig.~\ref{fig:Fig2}, upper panel; left column) models in excellent agreement with the DFT bands.  We study the electronic behavior in the crystal using the constructed four-band Wannier model.
Projecting the DFT Hamiltonian onto Wannier states yields a minimal model containing four orbitals in each unit cell $\psi_{\nu}$ with $\nu \in \{ 1,2,3,4 \}$. These are coupled in real space via four types of matrix elements: 
\begin{itemize}[noitemsep,topsep=0pt,leftmargin=*]
\item Intra-unit cell hopping $t_0 = 0.022$eV between $\psi_1$ and $\psi_2$ and also between $\psi_3$ and $\psi_4$.
\item Inter-unit cell hopping $t_1 = 0.028$eV from $\psi_1$ in unit cell $\vec{i}$ to $\psi_4$  in unit cell $\vec{i} + \vec{a}$ and  to $\psi_4$  in unit cell $\vec{i} + \vec{b} + \vec{c}$, and similarly from   $\psi_2$  in unit cell $\vec{i}$   to $\psi_3$  in unit cell $\vec{i} + \vec{a}$ and to $\psi_3$ in unit cell $\vec{i} + \vec{b} + \vec{c}$, and their Hermitian conjugates.
\item Inter-unit cell hopping $t_2 = 0.013$eV from $\psi_1$ in unit cell $\vec{i}$ to $\psi_3$ in unit cell $\vec{i} + \vec{b}$ and from  $\psi_2$ in unit cell $\vec{i}$ to $\psi_4$ in unit cell $\vec{i} + \vec{c}$, and their Hermitian conjugates.
\item Very small further-neighbor hoppings with magnitude $<0.01$eV that have a negligible effect on the physics which we ignore in the following, see Fig.~\ref{fig:Fig2}, upper panel; right column.
\end{itemize}

To better understand the  minimal model, we consider the $t_2=0$ limit (see Supplementary Information for more details). We contrast the electronic behavior of the model in this artificial $t_2=0$ limit against the nearly exact $t_2\neq 0$ limit in Fig.~\ref{fig:Fig2}, upper (right column) and lower panels.  Unlike in the $t_2 \neq 0$ case (Fig.~\ref{fig:Fig2}, lower panel; top), in the $t_2 = 0$ limit we find perfectly dispersionless bands along the two directions defined by the vectors orthogonal to the direction of $\vec{b}+\vec{c}-\vec{a}$ (Fig.~\ref{fig:Fig2}, lower panel; bottom). The band structure becomes flat  along the segments in $k$-space corresponding to these orthogonal directions (Fig.~\ref{fig:Fig2}, upper panel; right column). Thus, the vector $\vec{b}+\vec{c}-\vec{a}$ defines a special one-dimensional (1D) connectivity between Wannier orbitals in the minimal model. We can understand this 1D limit as follows. First setting $t_0 = 0$, we obtain two uncoupled 1D zigzag chains: one where an electron hops with amplitude $t_1$ from $\psi_1$ to $\psi_4$ one cell further in the $\vec{b} + \vec{c}$ direction, then again to $\psi_1$ one more cell further in the $-\vec{a}$ direction, and a second involving hopping between $\psi_2$  and $\psi_3$ in similar directions. Thus we see that electrons disperse along the $\vec{b}+\vec{c}-\vec{a}$ direction.  The zigzag pattern reflects the spatial orientation of the Wannier orbitals across different unit cells. A finite $t_0$ hopping amplitude couples the two zigzag chains via hopping between $\psi_1$ ($\psi_4$) on one chain and $\psi_2$ ($\psi_3$) on the other. This connectivity corresponds naturally to a ladder geometry with hopping $t_0$ along the rungs and with hopping $t_1$ along the legs, see Fig.~\ref{fig:Fig1}, lower panel. At $t_2 = 0$, the bands  are exactly particle-hole symmetric (Fig.~\ref{fig:Fig2}, lower panel; bottom) and exhibit a doubled Brillouin zone or, equivalently, a halved unit cell, as expected in a ladder structure whose primitive unit cell contains two orbitals. The coupled-ladder structure present in this system was not discussed in the terphenyl structures obtained theoretically in Ref.~\cite{JPhysChemCDFT}. 

We can now easily understand the $t_2\neq 0$ limit of the minimal model. We see that $t_2$ couples different legs of pairs of ladders, either in the $\vec{b}$ or in the $\vec{c}$ direction, see Fig.~\ref{fig:Fig1}, lower panel.   In this coupled-ladder construction,
an electron can hop for example from $\psi_1$ on one given ladder to $\psi_3$ on another in the $\vec{b}$ direction, and then move on the ladder arbitrarily until it arrives at  $\psi_4$, at which point it can hop to $\psi_2$ on a third ladder in the $-\vec{c}$ direction, or alternatively until it arrives at $\psi_2$, at which point it can hop to a different third ladder in $\vec{c}$ direction. The former process implies electrons disperse along the $\vec{b}-\vec{c}$ direction and the latter one tells us they also disperse along the $\vec{a}$ direction (the $y$-component of $\vec{b}+\vec{c}$).  This explains the full anisotropic electronic band dispersion in all three space and momentum directions shown in Fig.~\ref{fig:Fig2}. This 3D network of weakly coupled ladders must be  described by the full unit cell of four orbitals, giving rise to the enlarged zone in Fig.~\ref{fig:Fig2}, lower panel; top.

Making use of this information, we derive a Hamiltonian for this system in a more convenient coordinate system described by a body-centered cubic (BCC) lattice geometry. We introduce   $\vec{x}_{\mathrm{BCC}} = \vec{b} + \vec{c} - \vec{a}$,  $\vec{y}_{\mathrm{BCC}} = \vec{a}$ and $\vec{z}_{\mathrm{BCC}} = \vec{b} - \vec{c}$ (where $\vec{x}_{\mathrm{BCC}}\cdot \vec{y}_{\mathrm{BCC}} = \vec{y}_{\mathrm{BCC}}\cdot \vec{z}_{\mathrm{BCC}}= 0$), which define the principal axes along which electrons in the minimal model disperse.  We write primitive lattice vectors for the BCC lattice in terms of these vectors: $\vec{a}_1^{\rm P} = \frac{1}{2} (\vec{x}_{\mathrm{BCC}} + \vec{y}_{\mathrm{BCC}} + \vec{z}_{\mathrm{BCC}})$, $\vec{a}_2^{\rm P} = \frac{1}{2} (\vec{x}_{\mathrm{BCC}} - \vec{y}_{\mathrm{BCC}} + \vec{z}_{\mathrm{BCC}})$ and $\vec{a}_3^{\rm P} = \frac{1}{2} (\vec{x}_{\mathrm{BCC}} + \vec{y}_{\mathrm{BCC}} - \vec{z}_{\mathrm{BCC}})$.  In this coordinate system the minimal tight-binding model is given by
\begin{eqnarray}
&& \quad \hat{\mathcal{H}}_{\mathrm{t.b.}} = \hat{\mathcal{H}}_{\mathrm{L}} + \hat{\mathcal{H}}_{\mathrm{L\mbox{-}L}}\label{Eq:HBCC}, \\
\hat{\mathcal{H}}_{\mathrm{L}}  &=& -t_0 \sum_{\vec{i},\sigma}  \Big( \hatc^\dagger_{\vec{i},2,\sigma} \hatc_{\vec{i},1,\sigma}  + \hatc^\dagger_{\vec{i},4,\sigma} \hatc_{\vec{i},3,\sigma}  + {\rm h.c.} \Big) \nonumber \\
& &-t_1 \sum_{\vec{i},\sigma} \Bigg\{ \Big( \hatc^\dagger_{\vec{i}+\vec{a}_1^{\rm P}+\vec{a}_3^{\rm P},4,\sigma}  + \hatc^\dagger_{\vec{i}+\vec{a}_1^{\rm P}-\vec{a}_2^{\rm P},4,\sigma}
\Big) \hatc_{\vec{i},1,\sigma}  \nonumber \\
&& \quad \quad   + \Big( \hatc^\dagger_{\vec{i}+\vec{a}_1^{\rm P}+\vec{a}_3^{\rm P},3,\sigma}  + \hatc^\dagger_{\vec{i}+\vec{a}_1^{\rm P}-\vec{a}_2^{\rm P},3,\sigma}
\Big)  \hatc_{\vec{i},2,\sigma}   + {\rm h.c.}  \Bigg\}, \nonumber \\ \label{Eq:Hladder}  \\
\hat{\mathcal{H}}_{\mathrm{L\mbox{-}L}} &=&  -t_2 \sum_{\vec{i},\sigma}  \Big( \hatc^\dagger_{\vec{i}+\vec{a}_1^{\rm P},3,\sigma} \hatc_{\vec{i},1,\sigma}  +  \hatc^\dagger_{\vec{i}+\vec{a}_3^{\rm P},4,\sigma} \hatc_{\vec{i},2,\sigma}   + {\rm h.c.} \Big ), \nonumber \\ \label{Eq:HLL}  
\end{eqnarray}
where the operator $\hatc^\dagger_{\vec{j},\nu,\sigma}$ ($\hatc_{\vec{j},\nu,\sigma}$) creates (annihilates) a spin-$1/2$ electron with spin $\sigma \in \{\uparrow, \downarrow\}$ on $\psi_{\nu}$ with orbital index $\nu \in \{1,2,3,4\}$ located in a unit cell whose location is given by $\vec{j}$.
This Hamiltonian describes a system of 1D ladders coupled in a BCC lattice geometry in 3D space, as depicted in Fig.~\ref{fig:Fig1}, lower panel. The level of K doping found in this compound studied in experiment  corresponds to half filling of the electronic bands.  This type of model system of half-filled coupled ladders provides fertile ground for exploring emergent collective electronic and magnetic phases due to an interplay between correlations and dimensionality, as we discuss below.

{\em Correlations at strong coupling}.--- Given the large size of the molecular orbitals found in this system, it is reasonable to assume electron-electron correlations are predominantly local and proceed within a Hubbard~\cite{Hubbard1,Hubbard2} or Pariser-Parr-Pople~\cite{PPP1,PPP2} approach. We thus study correlation-driven physics using an interaction Hamiltonian that takes the form: $\hat{\mathcal{H}}_{\mathrm{t.b.}}+ U \sum_{\vec{i},\nu}  \hat{n}_{\vec{i},\nu\uparrow} \hat{n}_{\vec{i},\nu,\downarrow}$. Based on existing literature on aromatic compounds we  expect $U \sim 0.5$ $\mbox{-}$ $2\mathrm{eV}\gg t_0, t_1$ and $t_2$ for our system.  For these large $U/t$ (where $t \in \{t_0, t_1, t_2\}$) the electronic correlation opens a charge gap at half filling, inducing insulating behavior as also seen in experiment~\cite{ColumbiaTerphenyl}.  To unveil the nature of correlations in the half-filled state, we perform a strong-coupling expansion to leading order in ${t_0^2}/{U}$, ${t_1^2}/{U}$ and ${t_2^2}/{U}$, and obtain a magnetic spin  model with Heisenberg interactions:
\begin{eqnarray}
\hat{\mathcal{H}}_{\mathrm{t.b.};\mathrm{M}} &=& \hat{\mathcal{H}}_{\mathrm{L};\mathrm{M}}  + \hat{\mathcal{H}}_{\mathrm{L}\mbox{-}\mathrm{L};\mathrm{M}}, \label{Eq:HSC} \\
\hat{\mathcal{H}}_{\mathrm{L};\mathrm{M}}  &=& \sum_{\vec{i},\vec{j},\nu,\mu} \mathcal{J}_{\vec{i},\vec{j}}^{\nu,\mu} \Big( \vec{S}_{\vec{i}}^\nu \vec{S}_{\vec{j}}^\mu  - \frac{\hat{n}_{\vec{i},\nu} \hat{n}_{\vec{j},\mu}}{4} \Big), \label{Eq:HSCladder}  \\
\hat{\mathcal{H}}_{\mathrm{L}\mbox{-}\mathrm{L};\mathrm{M}} &=&  J_2 \sum_{\vec{i}}  \Big( \vec{S}_{\vec{i}}^1  \vec{S}_{\vec{i}+\vec{a}_1^{\rm P}}^3  - \frac{\hat{n}_{\vec{i},1} \hat{n}_{\vec{i}+\vec{a}_1^{\rm P},3}}{4} \nonumber \\
&& \quad \quad  \quad   + \vec{S}_{\vec{i}}^2  \vec{S}_{\vec{i}+\vec{a}_3^{\rm P}}^4  - \frac{\hat{n}_{\vec{i},2} \hat{n}_{\vec{i}+\vec{a}_3^{\rm P},4}}{4} \Big),\label{Eq:HLLSC}  
\end{eqnarray}
where  $\mathcal{J}_{\vec{i},\vec{j}}^{\nu,\mu} = \frac{1}{2} J_0 \delta_{\vec{i},\vec{j}} ( \delta_{\nu, 1} \delta_{\mu, 2} + \delta_{\nu, 2} \delta_{\mu, 1} + \delta_{\nu, 3} \delta_{\mu, 4} + \delta_{\nu, 4} \delta_{\mu, 3}) + \frac{1}{2} J_1 \big(\delta_{\vec{j}-\vec{i} , \vec{a}_1^{\rm P} +\vec{a}_3^{\rm P}} + \delta_{\vec{j}-\vec{i} , \vec{a}_2^{\rm P} +\vec{a}_3^{\rm P}}\big) (\delta_{\nu, 1} \delta_{\mu, 4} + \delta_{\nu, 4} \delta_{\mu, 1} + \delta_{\nu, 2} \delta_{\mu, 3} + \delta_{\nu, 3} \delta_{\mu, 2} )$ and  the magnetic moment of an electron in orbital $\nu$ at unit cell $\vec{j}$ is given by $\vec{S}_{\vec{j}}^\nu = \sum_{\alpha,\beta} \hatc^{\dagger}_{\vec{j},\nu,\alpha} \frac{\hbar}{2} \vec{\sigma}_{\alpha,\beta} \hatc_{\vec{j},\nu,\beta} $,  where $\vec{\sigma} = \{\sigma_{\mathcal{X}},\sigma_{\mathcal{Y}},\sigma_{\mathcal{Z}}\}$ is a vector of spin-$1/2$ Pauli matrices.  Here $J_0 = 4 t_0^2/U$, $J_1 = 4 t_1^2/U$ and $J_2 = 4 t_2^2/U$, where the $J_0$, $J_1$ and $J_2$ bonds  correspond, respectively, to the $t_0$, $t_1$ and $t_2$ bonds in Fig.~\ref{fig:Fig1}, lower panel. The couplings $J_0$, $J_1$ and $J_2$ respectively correspond to the couplings $J_b$, $J_a$ and $J_c$ obtained in Ref.~\cite{ColumbiaTerphenyl} by fitting energy differences of magnetically ordered states obtained in spin-polarized DFT calculations.  The ratios between the couplings are similar, and the reported magnitudes of $J_b$, $J_a$ and $J_c$  in Ref.~\cite{ColumbiaTerphenyl} would correspond to a $U\sim 0.4$eV.

{\em Magnetic behavior}.--- In the limit $J_2 = 0$, the system reduces to a set of uncoupled quasi-1D ladders, as discussed above.  The magnetic behavior of two-leg ladder systems is by now fairly well understood~\cite{StrongMillis-2legladder1,NoakWhiteScalapin-2legladder2,WhiteNoakScalapin-2legladder3,White-2legladder4,Tsvelik-2legladder5,Rev-2legladder6}. A ladder is formed of two coupled spin-$1/2$ chains, and thus supports a dimer phase formed of singlets localized over the ladder rungs. The ground state of this phase is a spin-gapped paramagnet. 
A non-zero value of $J_2$ couples the ladders, mixing spin-triplet states into the ground state, and a sufficiently large $J_2$ eventually induces magnetic ordering~\cite{DagottaCoupledLadders1,RiceCoupledLadders1,GiamarchiCoupledLadders1}. The value of $J_2$ required to induce a transition may exhibit non-monotonic behavior with temperature $T$ because the magnetically ordered state has gapless (spin wave) excitations and so has larger entropy than the spin-gapped paramagnetic state~\cite{Villain}, see the proposed phase diagram in Fig.~\ref{fig:Fig3}. At higher $T$ thermal fluctuations ultimately destroy ordering, and thus a larger $J_2$ is required to induce a transition~\cite{sachdev2011quantum}.
In our system $J_2/J_0 \approx J_2/J_1 \approx 1/4$,  which is likely sufficient to induce ordering into a 3D antiferromagnetic state at zero $T$, see the location of this compound on the phase diagram of the magnetic model shown in Fig.~\ref{fig:Fig3}. A precise determination of the exact behavior of this system in the  $T$-$J_2$ plane necessitates  development of new exact methods suitable for the study of coupled ladders in 3D space, a current challenge to theory and numerical approaches.  Studies of a few weakly coupled half-filled ladders in cylindrical geometries may be accessible to tensor network methods~\cite{DMRG}, a direction we leave to the future.

\begin{figure}[!]
\centering
\includegraphics[width=0.847\columnwidth]{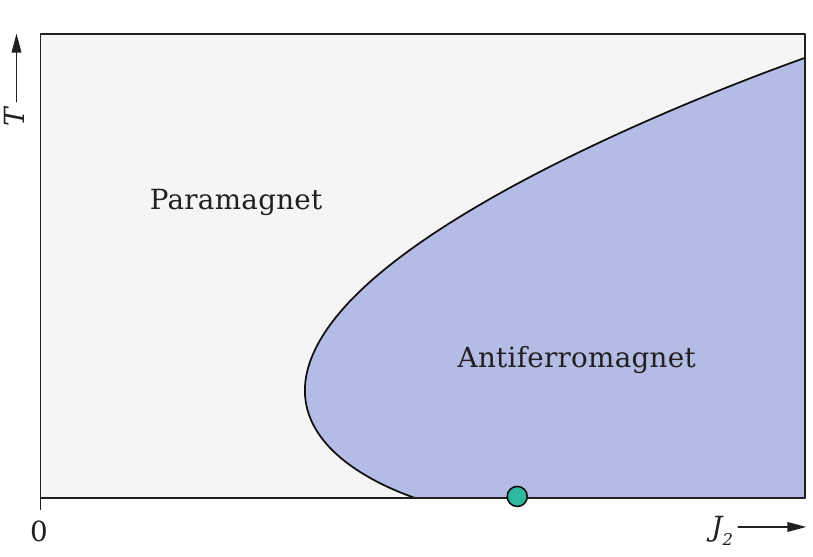}
\caption{{\bf Schematic phase diagram of coupled Heisenberg ladders in [K($\mathbf{222}$)]$_2$[$p$-terphenyl]$_3$ in the ladder-ladder coupling ($J_2$)-temperature ($T$) plane.} The dot denotes the proposed position of the compound studied in experiment~\cite{ColumbiaTerphenyl} on the phase diagram at $T=0$.}\label{fig:Fig3}
\vspace{-4mm} 
\end{figure}

{\em Light doping with holes}.--- The crystal can be lightly doped with holes via gating or by using an  oxidizing agent placed near the crystal surface. This situation gives rise to a $t$-$J_0$-$J_1$-$J_2$ model given by $\hat{\mathcal{H}}_{t\mbox{-}J_0\mbox{-}J_1\mbox{-}J_2} = \hat{\mathcal{P}}\Big[ \hat{\mathcal{H}}_{\mathrm{t.b.}} + \hat{\mathcal{H}}_{\mathrm{t.b.};\mathrm{M}} \Big]\hat{\mathcal{P}}$, where $\hat{\mathcal{P}}$ implements a constraint of no double occupancy that reflects the large onsite Coulomb repulsion of the strong-coupling limit.  Consider again  the $J_2 = 0$ limit.  Here, the problem reduces to the extensively studied doped $t$-$J$ ladder (with $J=J_0 \approx J_1$) in which the magnetic background mediates $d$-wave pairing of holes with concentration $\delta$, resulting in a phase with power-law superconducting correlations in the presence of a spin gap $\sim J_0$~\cite{Dopingladders1,Dopingladders2,Dopingladders3,Dopingladders4,DagottoReview}. In this 1D limit, the superfluid stiffness $\rho_{\rm s}$ is a fraction of $t  \delta$, where $t = t_0 \approx t_1$.  At sufficiently large hole concentrations, the superconducting phase gives way to regimes with Fermi liquid-like behavior~\cite{Dopingladders2}. For $J_2\neq 0$, the ladders are coupled, and the ultimate fate of the $d$-wave superconducting phase is an open problem whose solution in 2D and 3D space constitutes a major challenge to theory~\cite{DagottoScience,DwaveRev}.  In the absence of definitive results, we base our analysis on existing results for inhomogeneously coupled quasi-1D~\cite{WeaklyCoupledLaddersFtheory,Inhomogenous1,Inhomogenous2,Inhomogenous3,WeaklyCoupledLaddersDMRG} and finite-width (e.g. four-~\cite{4-legtJ} and eight-~\cite{4-legtJ} leg) ladders, and provide a qualitative discussion of the expected behavior in the $J_2\mbox{-}\delta$ plane  in Fig.~\ref{fig:Fig4}. We expect a superconducting region in the phase diagram, reflecting the survival of ladder $d$-wave  superconducting behavior for a range of optimal $\delta$ and $J_2$~\cite{WeaklyCoupledLaddersFtheory,WeaklyCoupledLaddersDMRG}.   This phase is stable for $T<T_{\rm c}$ and $T_{\rm c} \sim \rho_{\rm s}$ for small $\delta$.  From this we estimate that  this compound  may exhibit  superconductivity on hole-doping with a low $T_{\rm c}$ on the order of a few degrees Kelvin. A sufficiently large $J_2\sim J_0, J_1$, corresponding to homogeneously coupled ladders, or, a 3D network of spins, at zero $\delta$ gives rise to antiferromagnetic ordering, which becomes increasingly unstable to superconductivity and ultimately to Fermi liquid phases with increasing $\delta$, and thus a larger $J_2$ is needed in order stabilize the magnetic order.  From this we suggest that the parent anitferromagnetic compound obtained in recent experiments~\cite{ColumbiaTerphenyl} may give rise to superconductivity upon doping with holes for $\delta \sim 0.1$, see Fig.~\ref{fig:Fig4}.

 \begin{figure}[!b]
 \vspace{-2mm} 
\centering
\includegraphics[width=0.847\columnwidth]{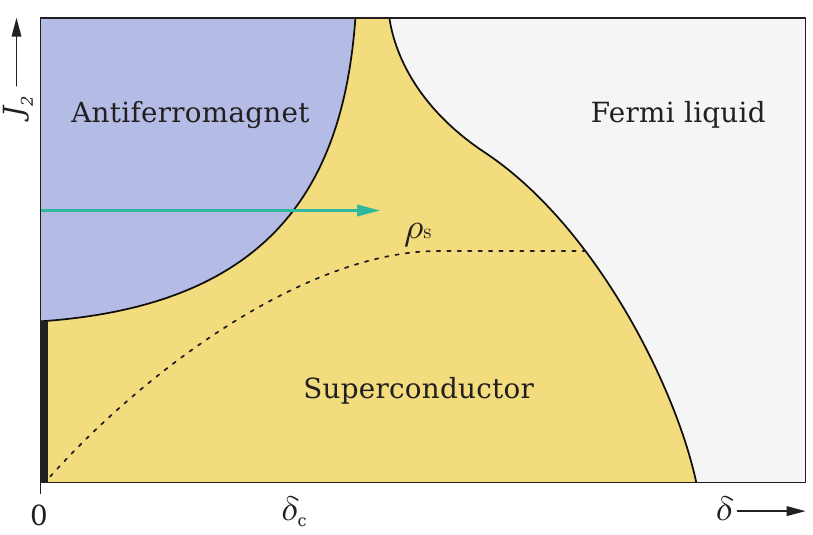}
\caption{{\bf Schematic $T=0$ phase diagram of  lightly doped coupled Heisenberg ladders in [K($\mathbf{222}$)]$_2$[$p$-terphenyl]$_3$ in the hole concentration ($\delta$)-ladder-ladder coupling ($J_2$) plane.} The  thick vertical line at $\delta=0$ denotes the non-superconducting spin-gapped zero doping phase, see Fig.~\ref{fig:Fig3}. The arrow depicts hole doping of the parent antiferromagnetic compound obtained in experiment~\cite{ColumbiaTerphenyl}, which becomes superconducting for $\delta\sim 0.1>\delta_{\rm c}$. We do not show here regions corresponding to charge-density wave order potentially emergent at much larger, commensurate dopings such as quarter filling~\cite{CommensurateCDW}.}\label{fig:Fig4}
\end{figure}

{\em Conclusions}.--- The K-doped terphenyl ladder materials open a new chapter in strongly correlated-electron behavior in the organics, potentially providing access to unconventional magnetic and superconducting phases. Recent studies~\cite{ChineseTerphenyl,ItalianTerphenyl} report temperature-dependent gaps in annealed terphenyl-based powders whose exact structural composition remains unsettled. Our theoretical analysis provides a clear picture of the many-body electronic band structure and correlations in the experimentally well-characterized, recently synthesized doped terphenyl crystals~\cite{ColumbiaTerphenyl}.  We show using DFT calculations and Wannier fitting that this complex material whose unit cell contains 444 atoms can be accurately described by a model of ladders sparsely coupled in 3D space with hopping matrix elements in the $\sim 10$ - $30 \mathrm{meV}$ range. We develop a strong-coupling approach in the limit of large Coulomb repulsion from which we argue that this half-filled system is likely an antiferromagnetic insulator. Upon hole doping this narrow-band system, we expect a regime of $d$-wave superconductivity whose signatures in experiment are of major interest. Because of the small values of the hopping matrix elements, ladder superconductivity in this system should occur at a low $T_{\rm c}$, on the order of a few degrees Kelvin.

From the perspective of materials science, understanding the difference in electronic structure of this system in relation to optimized doped terphenyl structures~\cite{JPhysChemCDFT} represents an important question for research.  On the theory side, the discovery of this ladder system calls for a systematic study of the physics of weakly coupled ladders in 3D space. Electron-phonon coupling is sometimes important in organic materials. Based on arguments for similar organic crystalline systems~\cite{e-ph1,e-ph2}, we expect that polaron (or at least small polaron) formation, and thus charge localization, to be unlikely, and that coupling to Peierls ~\cite{marchand2010sharp} phonon modes related to terphenyl-terphenyl bond distortions exerts the largest effect in transport, possibly leading to  unconventional electronic behavior~\cite{SousBipolaron}. A quantitative treatment of this problem however, requires  inclusion of the coupling of electrons to both Holstein~~\cite{holstein1959studiesI,holstein1959studiesII} and Peierls modes, a direction we plan to address in the future.

Future work will be devoted to a full theoretical characterization of the behavior of coupled ladder systems via tensor network and quantum Monte Carlo techniques in order to precisely pinpoint the doping range needed for inducing superconductivity, and to the experimental implementation of the conditions that will enable testing of the full range of predictions made in this work, thus opening a door to the observation of emergent electronic and magnetic phases in these novel and controllable organic crystalline systems.

\emph{Acknowledgements}.--- We acknowledge useful discussions with S. Kivelson, X. Roy, M.~L. Steigerwald and S. White. J. S. thanks A. B. Georgescu, J. Lee and especially J. Bonini for assistance with DFT calculations, and L. Reading-Ikkanda for assistance with creating schematics used in the figures. J.~S., N.~A.~G., C.~N., D.~R.~R. and A.~J.~M. acknowledge support from the National Science Foundation (NSF) Materials Research Science and Engineering Centers (MRSEC) program through Columbia University in the Center for Precision Assembly of Superstratic and Superatomic Solids under Grant No. DMR-1420634 and Grant No.  DMR-2011738.  C.~N. thanks Sheldon and Dorothea Buckler for their generous support.  J.~S. also acknowledges the hospitality of the Center for Computational Quantum Physics (CCQ) at the Flatiron Institute. The Flatiron Institute is a division of the Simons Foundation.

\providecommand{\noopsort}[1]{}\providecommand{\singleletter}[1]{#1}%

\clearpage

\break

\setcounter{equation}{0}

\title{
SUPPLEMENTARY INFORMATION\\
for\\
``Phonon-induced disorder in dynamics of optically pumped metals from\\ non-linear electron-phonon coupling''
}

\maketitle

\onecolumngrid

\pagenumbering{arabic}


\renewcommand{\thesection}{\Roman{section}}

\section{Crystal structure}

\paragraph{Chemical Stoichiometry.} We study the recently synthesized K-doped $p$-terphenyl compound~\cite{ColumbiaTerphenyl} with the chemical formula: [K($\mathbf{222}$)]$_2$[$p$-terphenyl]$_3$ corresponding to 2 K atoms, each within a $\mathbf{222}$ cryptand, for every 3 $p$-terphenyl (C$_{18}$H$_{14}$) molecules.  The cryptand is a polydentate ligand formed from 2 nitrogen, 6 oxygen, 18 carbon, and 36 hydrogen atoms with 62 atoms in total.  $p$-terphenyl consists of three linked benzene rings with a total of 32 atoms.  Each unit of [K($\mathbf{222}$)]$_2$[$p$-terphenyl]$_3$ thus consists of a total of 222 atoms.
\medskip

\paragraph{Primitive unit cell.} We use the pymatgen software library~\cite{pymatgen} to generate input files of the crystal data of our structure for density-functional theory (DFT) calculations, see details below.  We identify using pymatgen a primitive unit cell for the crystal with 444 atoms corresponding to two units of [K($\mathbf{222}$)]$_2$[$p$-terphenyl]$_3$ per unit cell, with cell parameters: $\vec{a} = \{0, 13.0237, 0.0\}$\si{\angstrom}, $\vec{b} = \{13.694074, 6.51185, 8.698852 \} $\si{\angstrom} and $\vec{c} = \{13.694074, 6.51185, -13.976448 \}$\si{\angstrom} in Cartesian coordinates. We obtain high-symmetry $k$ points in the Brillouin zone corresponding to this unit cell using pymatgen.
\medskip

\paragraph{Crystal symmetry.}  This system possesses a $C$2/$c$ monoclinic group lattice structure. As we show in the main text, the system's electronic structure is accurately described by a minimal tight-binding model with a connectivity that is isomorphic to one in a (distorted) body-centered cubic (BCC) lattice, see below.

\section{Methods}

\subsection{Density-functional theory (DFT) calculations}
We use the Quantum ESPRESSO softwards package~\cite{QuantumESSPRESSO} to perform density-functional theory (DFT) calculations within the generalized gradient approximation for the exchange-correlation functional parametrized by Perdew, Burke, and Ernzerhof (PBE)~\cite{PBE}, utilizing scalar relativistic ultrasoft pseudopotentials obtained from the PSlibrary~\cite{PSlibrary} for the atoms in the crystal. We use a kinetic energy cutoff of $60 \mathrm{Ry}$ for the wavefunctions and of $480 \mathrm{Ry}$ for the charge density and potential.  We employ the Marzari-Vanderbilt-DeVita-Payne ``cold smearing" approach~\cite{MVDP} for occupations with a Gaussian spreading of $1.5\times10^{-2}$.
We employ a convergence threshold for self-consistency of 
$10^{-9}$.  We self-consistently obtain electronic densities on a shifted Monkhrost-Pack 3 $\times$ 3 $\times$ 3 $k$-point grid, which we use to compute the electronic band structure along various $k$ paths suggested by pymatgen. We verify that a 3 $\times$ 3 $\times$ 3 Monkhrost-Pack grid suffices to achieve convergence of the interpolated band structure.

This system exhibits weak variation in the bond lengths between a certain subsets of pairs of $p$-terphenyl molecules along one particular orientation of the crystal, see Ref.~\cite{ColumbiaTerphenyl} for more details and discussions.  Here we use  in our calculations a structure with averaged bond lengths, ignoring the split short-long bond pattern.  Weak disorder effects we ignore will only lead to modest changes in the hopping parameters. In particular, disorder leads to a slight reduction in the hopping magnitude in the one-dimensional (1D) limit~\cite{ColumbiaTerphenyl}.
The cif file of this structure is available upon request.

 \subsection{Wannier localization}
 We use the Wannier90 software package ~\cite{Wannier90} to obtain a tight-binding representation of the DFT bands.  We first use
 the electron densities computed self-consistently in DFT calculations to generate in a non-self-consistent manner the densities at symmetry-equivalent $k$-points (this step inherits the same level of accuracy as that of the self-consistent calculation and incurs no greater error). We then use Wannier90  to obtain maximally localized Wannier orbitals and a tight-binding Hamiltonian from the $k$-space wavefunctions~\cite{WannierLocalization}.  We find excellent agreement between the DFT bands and the tight-binding fits of  four- and six-band (shown in the main text) models.

\section{Details of the interpretation of the electronic band structure}



Our approximate minimal tight-binding model in which we ignore matrix elements $<0.01\mathrm{eV}$ is given by the following Hamiltonian written in the Wannier basis:
\begin{eqnarray}
&& \hat{\mathcal{H}}_{\mathrm{t.b.}} = \hat{\mathcal{T}}_0  + \hat{\mathcal{T}}_1 + \hat{\mathcal{T}}_2, \label{Eq:Htb} \\ 
&\hat{\mathcal{T}}_0  &= -t_0 \sum_{\vec{i},\sigma}  \Big( \hatc^\dagger_{\vec{i},2,\sigma} \hatc_{\vec{i},1,\sigma}  + \hatc^\dagger_{\vec{i},4,\sigma} \hatc_{\vec{i},3,\sigma}  + {\rm h.c.} \Big), \label{Eq:HT0}\\
&\hat{\mathcal{T}}_1 & = -t_1 \sum_{\vec{i},\sigma} \Bigg\{ \Big( \hatc^\dagger_{\vec{i}+\vec{a},4,\sigma}  +
\hatc^\dagger_{\vec{i}+\vec{b}+\vec{c},4,\sigma} \Big) \hatc_{\vec{i},1,\sigma}  \nonumber \\
&& \quad \quad \quad \quad    +  \Big(\hatc^\dagger_{\vec{i}+\vec{a},3,\sigma} + \hatc^\dagger_{\vec{i}+\vec{b}+\vec{c},3,\sigma} \Big) \hatc_{\vec{i},2,\sigma}   + {\rm h.c.}  \Bigg\}, \label{Eq:HT1}  \\
&\hat{\mathcal{T}}_2  & =  -t_2 \sum_{\vec{i},\sigma}  \Big( \hatc^\dagger_{\vec{i}+\vec{b},3,\sigma} \hatc_{\vec{i},1,\sigma}    +   \hatc^\dagger_{\vec{i}+\vec{c},4,\sigma} \hatc_{\vec{i},2,\sigma}   + {\rm h.c.} \Big ),\label{Eq:HT2}  
\end{eqnarray}
where the operator $\hatc^\dagger_{\vec{j},\nu,\sigma}$ ($\hatc_{\vec{j},\nu,\sigma}$) creates (annihilates) a spin-$1/2$ electron with spin $\sigma \in \{\uparrow, \downarrow\}$ on a Wannier  orbital $\psi_{\nu}$ with orbital index $\nu \in \{1,2,3,4\}$ located in a unit cell whose location is given by $\vec{j}$, and $t_0 = 0.022\mathrm{eV}$, $t_1 = 0.028\mathrm{eV}$ and $t_2 = 0.013\mathrm{eV}$.

To understand the physics operative here we consider the hopping connecting orbitals across different unit cells. Consider the location of the Wannier orbitals within a unit cell given by $\vec{v}_{\psi_1} = \{6.127, 5.55, 2.223\}$\si{\angstrom}, $\vec{v}_{\psi_2} = \{7.567, 5.55, -4.862\}$\si{\angstrom}, $\vec{v}_{\psi_3} = \{-6.127, -5.55, -2.223\}$\si{\angstrom} and $\vec{v}_{\psi_4} = \{-7.567, -5.55, 4.862\}$\si{\angstrom}.  The large sizes of these Wannier orbitals imply that one must consider the specific orbital-orbital distance across different unit cells rather than cell-to-cell distance to understand the spatial structure of the network of hopping between orbitals in the crystal. The specific orientation in space of these orbitals reveals that within a unit cell there are a pair of orbitals $\{\psi_1, \psi_2\}$ related by inversion $x \rightarrow -x, y \rightarrow -y, z \rightarrow -z$ to $\{\psi_3, \psi_4\}$, and within a pair, orbitals share roughly the same $x$ and $y$ coordinates and are slightly displaced apart in $z$. We can thus visualize the large unit cell as composed of two symmetry-related pairs of orbitals, one pair in the $+y$ direction (which we later identify as $+y_{\mathrm{BCC}}$) contains $\{\psi_1, \psi_2\}$ with $\psi_1$ elevated in $z$ relative to $\psi_2$, and the other in the $- y$ direction  ($-y_{\mathrm{BCC}}$) contains  $\{\psi_3, \psi_4\}$ with $\psi_4$ elevated in $z$ relative to $\psi_3$. $\hat{\mathcal{T}}_0$ (Eq.~\eqref{Eq:HT0}) connects orbitals within a pair in the unit cell. To understand the network of hopping in $\hat{\mathcal{T}}_1$ (Eq.~\eqref{Eq:HT1}),  consider first the terms that mediate hopping of an electron between  $\psi_1$ and $\psi_4$
across different unit cells (first line of Eq.~\eqref{Eq:HT1}). $\hat{\mathcal{T}}_1$ contains a term that moves an electron from  $\psi_1$ at $\vec{i}$ with position $\{6.127, 5.55, 2.223\}$\si{\angstrom} to $\psi_4$ at $\vec{i}+\vec{a}$ in the neighboring cell with position  $\{-7.567, 7.4737, 4.862\}$\si{\angstrom}; this hopping predominantly moves the electron in the $-x$ direction. A second matrix element in $\hat{\mathcal{T}}_1$ connects $\psi_1$ at $\vec{i}$ with position $\{6.127, 5.55, 2.223\}$\si{\angstrom} to orbital $\psi_4$ at $\vec{i}+\vec{b} + \vec{c}$ with position $\{19.8211, 7.4737, -0.4156\}$\si{\angstrom}; this hopping predominantly moves the electron in the $+x$ direction. $\psi_4$ at $\vec{i}+\vec{a}$ and $\psi_4$ at $\vec{i}+\vec{b} + \vec{c}$ do in fact share the same $y$-coordinate and are only relatively slightly apart in $z$. Since  $\psi_1$ and $\psi_4$ are symmetry related to $\psi_2$ and $\psi_3$, we can understand the terms in $\hat{\mathcal{T}}_1$ mixing $\psi_2$ and $\psi_3$ in a similar fashion.
The 1D limit is now apparent.  We observe a zigzag pattern of connections between $\psi_1$ and $\psi_4$ ($\psi_2$ and $\psi_3$) across different unit cells and the two zigzag patterns intersect.  These connections due to  $\hat{\mathcal{T}}_0$ (Eq.~\eqref{Eq:HT0}) and $\hat{\mathcal{T}}_1$ (Eq.~\eqref{Eq:HT1}) in the $t_2 = 0$ limit map perfectly onto a network described by a ladder geometry with hopping of amplitude $t_0$ within a rung between $\psi_1$ and $\psi_2$ and similarly between $\psi_3$ and $\psi_4$ and hopping of amplitude $t_1$ along the legs between $\psi_1$ and $\psi_4$ and identically between $\psi_2$ and $\psi_3$, see Fig.~1, lower panel.  This lattice geometry has a primitive unit cell of two orbitals, explaining the doubled spectrum in $k$-space.

$t_2\neq 0$ results in ladder-ladder hopping in all three space directions. To unveil the orientation of couplings between ladders, we consider the hopping terms in $\hat{\mathcal{T}}_2$ (Eq.~\eqref{Eq:HT2}).  One term in $\hat{\mathcal{T}}_2$ couples $\psi_1$ in unit cell $\vec{i}$ at $\{6.127, 5.55, 2.223\}$\si{\angstrom} to $\psi_3$ in unit cell $\vec{i}+ \vec{b}$ at $\{7.5671, 0.9619, 6.4759\}$\si{\angstrom}; this hopping  moves an electron predominantly in both the $-y$ and $+z$ directions from one leg of a given ladder to a different leg of a second ladder. The other term in $\hat{\mathcal{T}}_2$ couples $\psi_2$ in unit cell $\vec{i}$ at $\{7.567, 5.55, -4.862\}$\si{\angstrom} to $\psi_4$ in unit cell $\vec{i}+ \vec{c}$ at $\{6.1271, 0.9619, -9.1144\}$\si{\angstrom}; this hopping moves an electron predominantly in both the $-y$ and $-z$ directions from the second leg of the original ladder to the first of a third ladder.  We see that the original ladder is sparsely coupled to two others, both located at about the same position in $x$ and $y$ but apart in $z$, see Fig.~1, lower panel. Since different legs of the different ladders are coupled by $t_2$, $\hat{\mathcal{H}}_{\mathrm{t.b.}}$ in Eq.~\eqref{Eq:Htb} eventually allows electrons to traverse the network of all ladders spanning 3D space, giving rise to full, albeit anisotropic, dispersion in all three space/ momentum directions, as shown in Fig.~2. This orientation of couplings between ladders actually reflects a distorted body-centered cubic (BCC) lattice geometry (see Fig.~1, lower panel for a sketch of an idealized BCC structure of this system) in which we can now recast the minimal tight-binding model, see main text for details and extended discussion.

\end{document}